\documentclass[%
 reprint,
 superscriptaddress,
 amsmath,amssymb,
 aps,
 twocolumn,
]{wlscirep}

\usepackage[english]{babel}
\usepackage{url}
\usepackage{graphicx}
\usepackage{dcolumn}
\usepackage{bm}
\usepackage{amsmath}
\usepackage{hyperref}
\usepackage{comment}

\usepackage{physics}

\usepackage[autostyle]{csquotes}
\MakeOuterQuote{"}

\usepackage{xcolor}
\definecolor{hamza-color}{RGB}{0,80,250}
\newcommand{\revision}[1]{{ #1 }} 
\newcommand{\revisiontwo}[1]{{ #1 }}

\begin{document}







\title{Spin-optomechanical cavity interfaces by deep subwavelength phonon-photon confinement}

\author[1,*]{Hamza Raniwala}
\affil[1]{Department of Electrical Engineering and Computer Science, Massachusetts Institute of Technology, Cambridge, MA 02139, USA}

\author[1]{Pratyush Anand}

\author[2]{Stefan Krastanov}
\affil[2]{University of Massachusetts Amherst, Amherst, MA, USA}

\author[3]{Matt Eichenfield}
\affil[3]{Sandia National Laboratories, Albuqurque, NM, USA}

\author[1,4]{Matthew Trusheim}
\affil[4]{Army Research Laboratory, Massachusetts Institute of Technology, Cambridge, MA 02139, USA}

\author[1,**]{Dirk R. Englund} 

\affil[*]{raniwala@mit.edu}
\affil[**]{englund@mit.edu}


\date{\today}

\begin{abstract}








A central goal of quantum information science is the transfer of qubits between space, time, and modality. Spin-based systems in solids have emerged as leading quantum memories, but high-fidelity transfer of their quantum states to telecom optical fields remains challenging. Here, we introduce an efficient phonon-mediated interface between spins in a 1D diamond nanobeam optomechanical crystal and telecom optical fields by a simultaneous deep-subwavelength confinement of optical and acoustic fields with mode volumes $V_\mathrm{mech}/\Lambda_\mathrm{p}^3\sim 10^{-5}$ and $V_\mathrm{opt}/\lambda^3\sim 10^{-3} $, respectively.
We show that subwavelength phonon confinement boosts the spin-mechanical coupling rate of Group IV silicon vacancy (SiV$^-$) centers by an order of magnitude to \revisiontwo{$\sim 32$ MHz}, while retaining high acousto-optical couplings. The engineered optical cavity couples to the atomic spin ground state irrespective of the spin's native excited states, avoiding spectral diffusion.
Using Quantum Monte Carlo simulations, we estimate entanglement fidelities exceeding 0.96 between two such remote spin-optomechanical interfaces mediated by a heralded telecom optical field. Since this interface decouples the spin system's excited states from the photonic bus, we anticipate broad utility beyond diamond emitter-telecom systems to most solid state quantum memories.

\end{abstract}

\maketitle

\section{Introduction}
The interaction of light with solid matter via radiation pressure forces is a remarkable phenomenon whose discovery dates back to the 17th century~\cite{Kepler, aspelmeyer2014cavitybook}. In recent decades, progress on understanding and engineering this light-matter interaction has produced groundbreaking experiments in cavity optomechanics, including laser feedback cooling~\cite{ashkin1978trapping}, parametric light-matter processes in kg-scale~\cite{cuthbertson1996parametric} and picogram-scale~\cite{eichenfield2009optomechanical, eichenfield2009picogram,chan2012optimized} optomechanical systems, and laser cooling of mechanical modes to their ground state~\cite{wilson2007theory, chan2012optimized}. These quantum optics-like experiments have paved the way for optomechanical devices to be used in quantum transduction~\cite{vainsencher2016bi, mirhosseini2020quantum, forsch2020microwave, jiang2020efficient, wu2020microwave} and entanglement~\cite{riedinger2018remote, zhong2020entanglement}.

Solid-state vacancy-defect complexes are a developing technology that is complementary to cavity optomechanics. These complexes are atomic defects in dielectric media, such as diamond, that can be intentionally created in a dielectric lattice~\cite{childress2007coherent, hepp2014electronic}. The free electron spin or nuclear spin of the resulting lattice vacancies can be coherently controlled as solid state quantum bits.\revision{~\cite{childress2007coherent, wolfowicz2021quantum}}. Additionally, research efforts demonstrating acoustic control of spin centers has opened the door to multi-modality quantum systems, such as spin-optomechanical interfaces \cite{ruskov2013onchip,maity2020coherent,shandilya2021optomechanical}. These complex coupled systems can potentially allow for dark-state operation of spin centers, optical-to-spin quantum transduction, and new architectures for quantum repeaters in a quantum network.

\begin{figure*}[ht]
    \centering
    \includegraphics{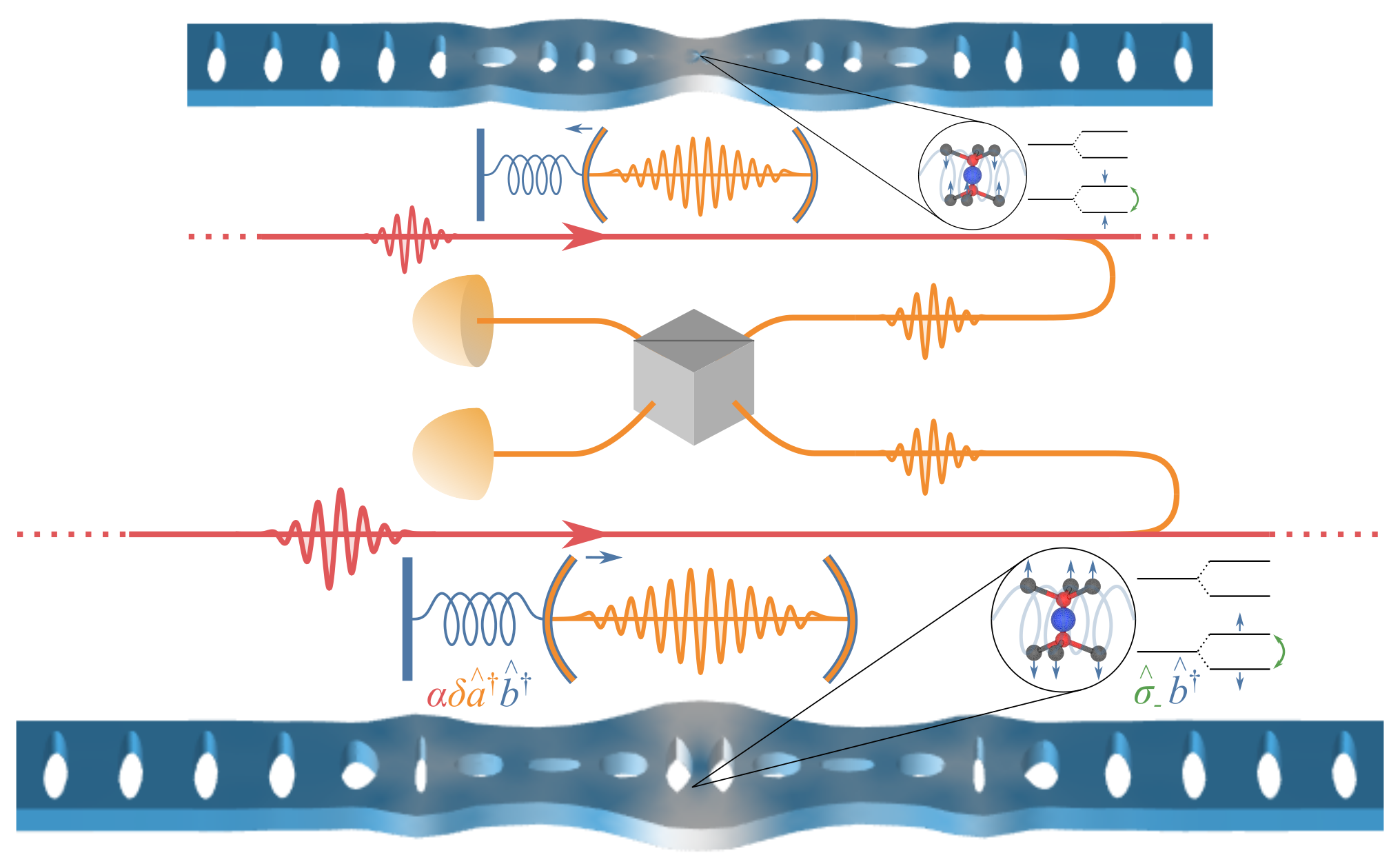}
    \caption{Depiction of spin entanglement via a spin-optomechanical interface. Each node contains an optical resonator (orange cavity / operator) coupled to a mechanical resonator (blue spring / operator), with an embedded color center (inset / green operator). A pump (red) is used to induce a two-mode squeezing in the opto-mechanical system. The leakage of an optical photon (orange waveguide) and its detection (gray detector) herald the creation of a single mechanical phonon. A beamsplitter (in gray) can be used to "erase" the knowledge of which is the original source of the photon, leading to the heralding of an entangled state $|10\rangle\pm|01\rangle$ between two neighboring nodes. The phase depends on which of the two detectors clicked.}
    \label{fig:repeaters}
\end{figure*}

Here, we propose an ultra-small mechanical and optical mode volume spin-optomechanical interface in diamond for strong coupling between the mechanical mode of an optomechanical resonator and an embedded group IV defect-vacancy complex. Our device introduces an optical resonance to ultra-small mechanical cavities previously used for spin interfacing~\cite{schmidt2020acoustic}. Critically, our device dramatically reduces mechanical mode volumes \revision{This design is predicated on a central tapering method used in photonic crystals and similar devices \cite{robinson2005ultrasmall, choi2017self, hu2018experimental, bozkurt2022deep}.} We show that this device can be used to interact with a vacancy without optically exciting the spin at its native wavelength, operating at the cavity wavelength instead through a optomechanically mediated interaction. Hence, using quantum modeling, we explore utilizing this spin-optomechanical interface for entanglement protocols in quantum networks in a setup depicted in Fig. \ref{fig:repeaters}. We use the modeled entanglement rates to feedback on a computer-aided design of an optimized spin-optomechanical interface.

\begin{figure*}[hbt!]
    \centering
    \includegraphics[width=\textwidth]{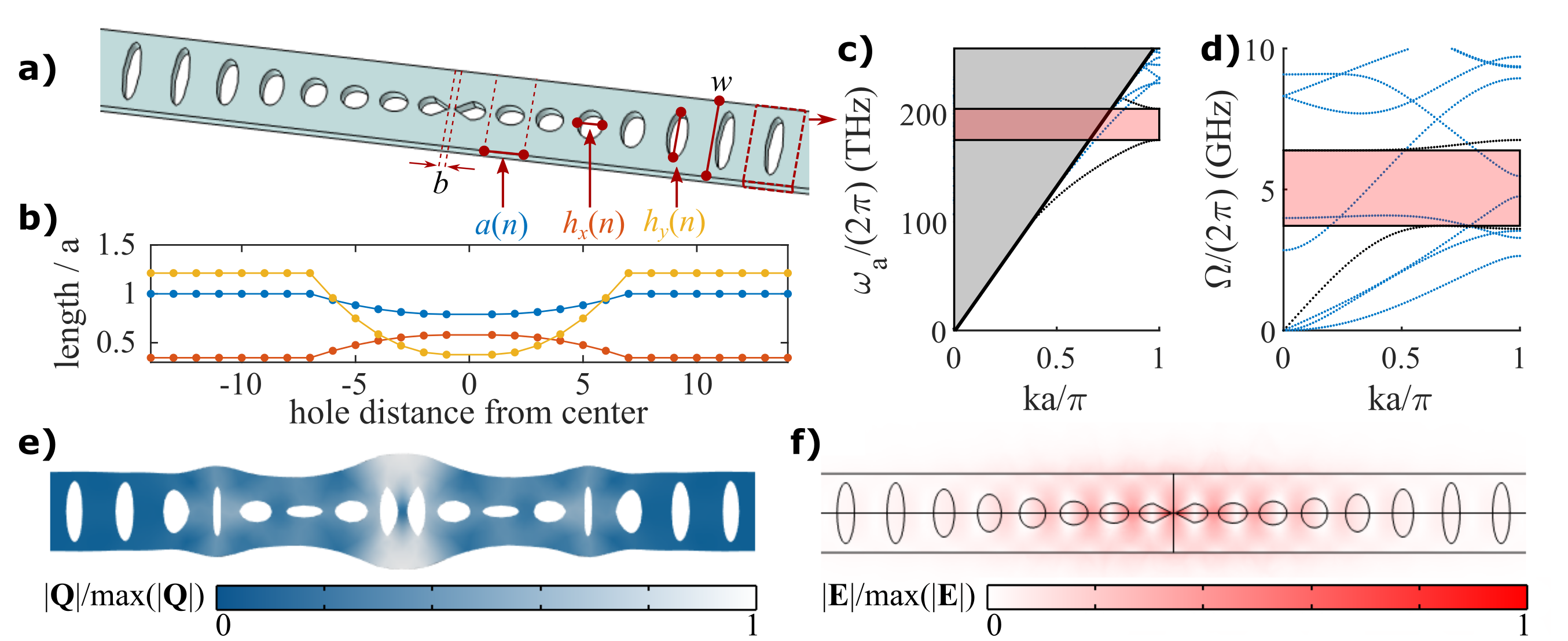}
    \caption{Diamond rectangular 1D nanobeam OMC with embedded concentrator, drawing from previous examples in silicon~\cite{eichenfield2009optomechanical,chan2012optimized} and diamond~\cite{burek2016diamond,cady2019diamond,maity2020coherent} as well as ultrasmall mode volume photonic and phononic crystals~\cite{choi2017self,schmidt2020acoustic}. (a) Diagram of the nanobeam photonic crystal. Free parameters include taper width $b$; unit cell period as a function of cell number $n$, $a(n)$; unit ellipse width $h_x(n)$ and height $h_y(n)$; and beam width $w$ alongside beam thickness $t$. (b) Plot of quadratically varying $a(n)$, $h_x(n)$, and $h_y(n)$ on either side of the beam center. This characterizes the cavity parametrized by Table \ref{optomech table} later in the text.
    (c) optical and (d) mechanical bandstructure for the mirror unit cell of the cavity, providing a 28.7 THz bandgap around and a 2.41 GHz mechanical bandgap. (e) mechanical displacement and (f) electric field norm profiles of the \revisiontwo{5.39 GHz mechanical mode and 200.2 THz} optical mode of the cavity. These simulations are for parameters $\{h_{y_d},h_{x_d},a_{d}\} = \{218.2,334.8,456.56\}$ nm, $\alpha = 135^{\circ}$, rest of the parameters are same as from Table~\ref{optomech table}.}
    \label{fig:bandstructures}
\end{figure*}

\section{Results}
\subsection{Theory of Spin-Optomechanical Coupling}

A spin-optomechanical interface accomplishes two effects. First, it couples the photonic mode of a photonic crystal cavity to the phononic modes of the crystal in a pump-driven interaction. Next, it couples the spin transition of a solid-state color center to the same phononic modes. Let us denote the operating frequency of the photonic mode as $\omega_\mathrm{a}$, the spin transition frequency as $\omega_{\sigma}$, and the pump beam frequency as $\omega_\mathrm{p}$. Without loss of generality, we assume only a single phononic mode $\Omega$ is nearly resonant with the pump detuning, such that $\Delta = \omega_\mathrm{p} - \omega_\mathrm{a} \approx \Omega$. Then we can simplify the system Hamiltonian by considering only a single phononic mode. In this picture, the unperturbed Hamiltonian $\hat{H}_0$ can be written as
\begin{equation}
    \hat{H}_0 = \hbar \omega_\mathrm{a} \hat{a}^\dag\hat{a} + \hbar\Omega\hat{b}^\dag\hat{b} + \frac{\hbar\omega_{\sigma}}{2}\hat{\sigma}_\mathrm{z}.\label{H0}
\end{equation}

Here, $\hat{a}^\dag,\hat{a}$ and $\hat{b}^\dag,\hat{b}$ are the ladder operators of the photonic and phononic modes, respectively, and $\hat{\sigma}_j$ is the spin qubit's $j$-Pauli operator. 


Additionally, the parametric coupling between the mechanical and optical resonators takes the form $\hat{H}_\mathrm{om} = \hbar g_\mathrm{om}\hat{a}^\dag\hat{a}\left(\hat{b}^\dag + \hat{b}\right)$, i.e., an optical resonance shift dependent on the position of the mechanical resonator.  To linearize this interaction, we drive the optical cavity with a pump $\omega_\mathrm{p} = \omega_\mathrm{a} + \Delta$.
By applying the rotating wave approximation and rewriting the photon ladder operators around a mean population $\overline{a}$ as $\hat{a} \rightarrow \overline{a} + \hat{a}$, \revisiontwo{we arrive at the typical optomechanical interaction Hamiltonian in the blue-detuned regime,

\begin{equation}
    \revision{\hat{H}_\mathrm{om-bl} = \hbar g_\mathrm{om}\overline{a}}\left( \hat{a}^\dag \hat{b}^\dag + \hat{a}\hat{b}\right).\label{Hom-bl}
\end{equation}

In the red-detuned regime, we get the following Hamiltonian
\begin{equation}
    \revision{\hat{H}_\mathrm{om-rd} = \hbar g_\mathrm{om}\overline{a}}\left( \hat{a}^\dag \hat{b} + \hat{a}\hat{b}^\dag\right).\label{Hom-rd}
\end{equation}
}
Next we consider the spin-mechanical interaction. In a spin-strain interaction picture, this is generated by deformation-induced strain causing a level shift in the spin qubit transition energy. This level shift is described by the spin-mechanical interaction Hamiltonian
\begin{equation}
    \hat{H}_\mathrm{sm} = \hbar g_\mathrm{sm}\left(\hat{\sigma}_{+}\hat{b} + \hat{\sigma}_{-}\hat{b}^\dag\right).\label{Hsm}
\end{equation}

Here, $g_\mathrm{sm}$ is the strain-induced coupling by the zero-point fluctuation of the mechanical resonator and $\hat{\sigma_{\pm}} = \frac{1}{\sqrt{2}}\left(\sigma_\mathrm{x} \pm i\sigma_\mathrm{y}\right)$. As such, any phonon excitation will induce zero-point coupling between the spin qubit and resonator phonon and vice versa. 

An efficient spin-optomechanical interface requires tuning of couplings $g_\mathrm{om}$ and $g_\mathrm{sm}$ as well as quality factors $Q_\mathrm{opt}$ and $Q_\mathrm{mech}$ for a targeted experiment. We delve into the design considerations that affect these parameters below.

\subsection{Device Simulations}
\begin{figure*}[t]
    \centering
    \includegraphics{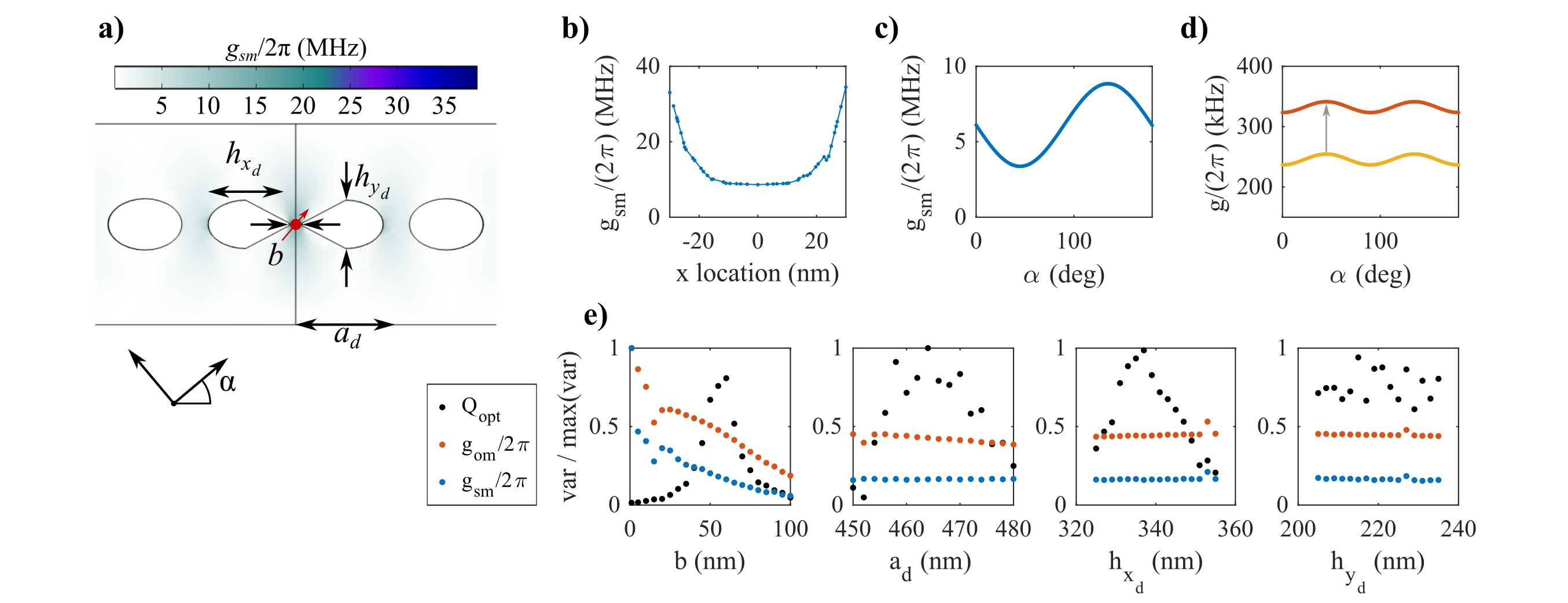}
    \caption{Analysis of spin-mechanical coupling profiles and the defect cell parameters that affect this coupling. (a) and (b) show the 2D and 1D spin-mechanical coupling $g_{sm}$ profile of the spin-optomechanical crystal breathing mode, with expected position of the Group IV spin overlayed. The parameters $b$, $a_d$, $h_{x_d}$, and $h_{y_d}$ impact the spin-mechanical coupling strength among other variables. The one-dimensional profile in the center of the bridge shows that strain is maximized at the bridge edge but retained in the center away from sidewalls. The crystal axis orientation in the device affects $g_{sm}$ (c) and $g_{om}$ (d, red) by changing the strain elements the spin sees and the photoelastic effect for $g_{pe}$ (d, yellow), which sums with a constant $g_{mb}$ (d, gray). Sweeping $b$, $a_d$, $h_{x_d}$, and $h_{y_d}$ about a naive parameter set shows that the parameters governing device performance, namely $g_{sm}$, $g_{om}$, and $Q_{opt}$ are maximized about different variables, motivating protocol-based numerical optimization of the exact design parameters. $Q_{mech}$ was routinely near $10^6$ or greater in the sweeps and not depicted.  These sweeps are centered around the following parameters $\{h_{y_d},h_{x_d},a_{d}\} = \{220.5,341.25,456.75\}$ nm, $\alpha = 135^{\circ}$, rest of the parameters are same as from Table~\ref{optomech table}.}
    \label{fig: sweeps}
\end{figure*}

At the core of our proposal is a strain concentrator embedded in a  one-dimensional optomechanical crystal (1D OMC) \revision{with rectangular cross-section}(Fig.\ref{fig:bandstructures}a).
The 1D OMC consists of a nanobeam with periodically etched ellipses, $2 n_d$ of which are adiabatically morphed into a defect cell. We then modify the central defect cells by tapering to a width $b$ using a linear taper. \revisiontwo{We simulate (at $b=60$ nm, $n_{d}=6$) an optical mode of frequency $\omega_\mathrm{a}/(2\pi) \approx 200.2$ THz and $Q_\mathrm{opt} \approx 9.6\times 10^4$ (Fig.~\ref{fig:bandstructures}(f)), which lies in the mirror cells' 28.7 THz optical bandgap from 175.28 THz to 203.98 THz (Fig.~\ref{fig:bandstructures}(b)). We predict an acoustic resonance around $\Omega = 5.39$ GHz (Fig.~\ref{fig:bandstructures}(e)) between the 2.41 GHz acoustic bandgap from 4.96 GHz to 7.37 GHz (Fig.~\ref{fig:bandstructures}(d))}. 

We note here that the parameters governing the defect cell \{$b$, $a_{d}$, $h_{x_d}$, $h_{y_d}$\}, as well as the crystal orientation in the device $\alpha$, heavily affect \{$g_{sm}$, $g_{om}$, $Q_{opt}$\} which are critical to the device performance ($Q_{mech}$ is relatively unaffected by the device geometry as long as the phonon frequency $\Omega$ lies within the acoustic bandgap, and is rather limited by phonon-thermally and materially-governed phonon-phonon processes ~\cite{akhiezer1939on,landau1939absorption,woodruff1961absorption,tabrizian2009effect}). As such, we explore the behavior of these performance parameters as a function of the defect unit cell parameters in Fig. $\ref{fig: sweeps}$. From these sweeps, we find a mix of simple and non-trivial relations between the defect cell parameters and the performance parameters. Setting $\alpha = \frac{3\pi}{4}$ rad, for example, maximizes both $g_{om}$ and $g_{sm}$ by simultaneously maximizing the photoelastic contribution to $g_{om}$--$g_{pe}$--and the strain along the transverse axis of an emitter in the center of the spin-optomechanical interface. 

Additionally, decreasing $b$, which can be thought of as the spring constant in the central bridge, increases the strain energy density, or equivalently the mechanical mode volume \cite{schmidt2020acoustic}, in the central bridge of the spin-optomechanical interface. We estimate through FEM that $V_\mathrm{mech}/\Lambda_\mathrm{p}^3$ and $V_\mathrm{mech}/\Lambda_\mathrm{s}^3$ drop from $\sim 10^{-4}$ and $\sim 10^{-3}$, respectively, to $\sim 10^{-6}$ and $\sim 10^{-5}$, respectively, as $b$ decreases from 100 nm to 20 nm. Here, $\Lambda_\mathrm{p}$ and $\Lambda_\mathrm{s}$ are the longitudinal and shear \revision{wavelengths} in bulk diamond~\cite{schmidt2020acoustic}. As $V_\mathrm{mech}$ decreases, $g_\mathrm{sm}$ increases, which also increases the "mechanical Purcell enhancement." $V_\mathrm{opt}/\lambda^3$ or $V_\mathrm{opt}/(\lambda/n)^3$ similarly decrease from $\sim 10^{-2}$ and $\sim 10^{-1}$, respectively, to \revisiontwo{$\sim 10^{-3}$ and $\sim 10^{-2}$}, respectively, with decreasing $b$--a beneficial effect for simultaneoxusly concentrating the cavity mechanical and optical modes. Here, $\lambda$ is the free space cavity wavelength, and $n$ is the refractive index of diamond. Practically, we find that both $g_{sm}$ and $g_{om}$ increase as $b$ is made as small as possible, limited by fabrication constraints.

However, the effects of \{$a_d$, $h_{x_d}$, and $h_{y_d}$\} on the performance parameters are less predictable, owing to a possible interplay between these parameters in the defect cell geometry. In previous studies on optomechanical crystals, these parameters are numerically optimized to yield the best performance parameter set \cite{chan2012optimized}. In this study, which considers a complicated tripartite interface of an optical cavity, a phononic cavity, and an embedded spin system, we would like to motivate the optimization not just by \{$g_{sm}$, $g_{om}$, $Q_{opt}$, $Q_{mech}$\} but by a protocol that utilizes this interface. We consider non-classical spin state heralding and opto-mechanically mediated entanglement between two spins in the sections below.

\subsection{Spin-Mediated Entanglement via DLCZ}

The controlled opto-mechanical two-mode squeezing represented by Eq.~\ref{Hom-bl} enables us to herald the creation of single phonons in the mechanical resonator. Such excitations can then be deterministically transferred to the spin via the spin-mechanical interaction in Eq. ~\ref{Hom-rd} for long term storage. Crucially, if we employ the Duan, Lukin, Cirac, and Zoller's ~\cite{duan2001long,krastanov2021optically} entangling protocol, we can herald an entangled $|01\rangle\pm|10\rangle$ state in two remote mechanical resonators. Each of the two mechanical resonators can then deterministically swap its content with their embedded spins\revision{ by DC strain-actuating the optomechanical crystal to detune the spin from the mechanical mode (see Supplement)}, leading to two remote entangled long-lived spins for use in quantum networking.


The DLCZ protocol is, at its core, two single-phonon heralding experiments running in parallel as seen in Fig.~\ref{fig:repeaters}. However, the detector triggering the heralding is placed after a "path erasure apparatus", e.g. a simple 50-50 beamsplitter. Therefore, when a phonon is heralded by the detection of a photon, the phonon is in equal superposition of being in the left or in the right node. This results in the two mechanical resonators being in the state $|01\rangle\pm|10\rangle$, with the phase depending on which detector clicked. For details on this path erasure consult \cite{duan2001long,krastanov2021opticallyheralded}.

\begin{figure}[bt]
    \centering
    \includegraphics[width=\columnwidth]{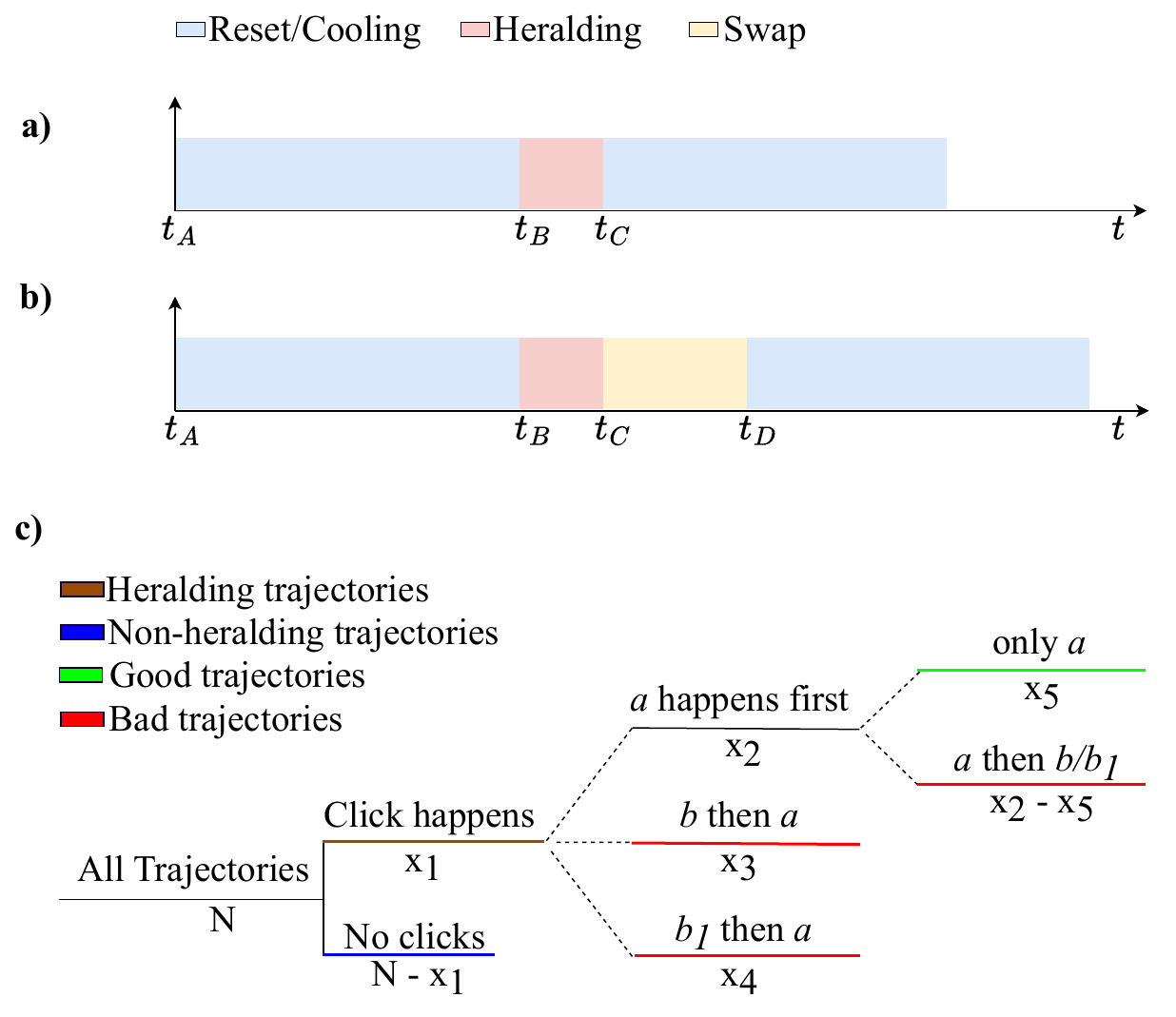}
    \caption{Schematic of the full protocol depending on the (a) absence or (b) presence of a photon click on the photo-detectors at $t=t_{C}$. (c) Map of all possible quantum trajectories. We start with N trajectories. Out of them, there is a photon click in $x_1$ of them. After getting a click, there are three possibilities, where events $a$, $b$ and $b_1$ correspond to the three collapse operators described in section 2.3. The spin-phonon swap is performed on all the brown trajectories, but ideally only the green ones are the good trajectories, which means that the red ones, lead to the infidelity of our heralding plus swap protocol. Using the trajectory approach, we can easily estimate, $P_{herald}$ = $\frac{x_1}{N}$ and $p_{good-traj}$ = $\frac{x_5}{x_1}$.}
    \label{fig:protocol}
\end{figure}

Hence, we develop a protocol for single-spin heralding by first initializing our spin-phonon system in the ground state via optomechanical cooling and then performing single-phonon heralding and swapping using optomechanical squeezing and spin-swapping. The protocol is shown in Fig. ~\ref{fig:protocol}a-b, where we (1) cool, (2) herald on the creation of an entangled phonon, and (3) conditionally swap to the vacancy spin state.

\revisiontwo{We begin by red-sideband optomechanical cooling of the phonon mode with the spin-mechanical coupling on for time $t_{B}-t_{A}$. This cools the phonon and coupled spin modes to their ground state. Then we utilize the blue-detuned Hamiltonian to perform the heralding protocol for a time $t_{C}-t_{B}$ shorter than the time for the spin or optomechanical cavities to thermally repopulate. After the heralding, the controlled swap operation is implemented depending on if we get a photon click. Then the full system Hamiltonian is
\begin{equation}
    \hat{H}_{cool} = \hat{H}_0 + \hat{H}_\mathrm{om-rd} + \hat{H}_\mathrm{sm}.\label{Hcool}
\end{equation}
\begin{equation}
    \hat{H}_{herald} = \hat{H}_0 + \hat{H}_\mathrm{om-bl} + \hat{H}_\mathrm{sm}.\label{Hherald}
\end{equation}
\begin{equation}
    \hat{H}_{swap} = \hat{H}_0 + \hat{H}_\mathrm{sm}.\label{Hswap}
\end{equation}
}

Below we study the fidelity and success probability of the single-phonon heralding protocol, as its performance directly affects the performance of the overall entanglement protocol. In this process, when writing down kets, we will use the Fock basis of the optical and mechanical modes, written down in that order, e.g., $|01\rangle$ is zero photons and one phonon. Two processes are involved in the single-phonon heralding: the two-mode squeezing in Eq.~\ref{Hherald} which leads to the mapping $|00\rangle\rightarrow|00\rangle+\varepsilon|11\rangle+\mathcal{O}(\varepsilon^2)$; and the leakage into a waveguide and subsequent detection of the photon, which projects on the $\varepsilon|11\rangle+\mathcal{O}(\varepsilon^2)$ branches. To properly derive the dynamics, we will use a stochastic master equation and we will track the most-probable quantum trajectories manually. The dynamics is governed by the equation
\begin{equation}
    \hat{H}_\textrm{stoch} = \hat{H}_0 + \hat{H}_\mathrm{om-bl} + \hat{H}_\mathrm{sm} - \frac{i}{2}\sum_{c \in \{a,b,b^\dag\}} \gamma_c\hat{c}^\dag\hat{c},
\end{equation}
where the the sum over jump operators $\hat{c}$ provides a way to track the chance for discontinuous jumps. If $\ket{\psi(t)}$ is the state obtained after evolving $\ket{00}$ under $\hat{H}_\textrm{stoch}$, then the probability density for a jump $\hat{c}$ is $\textrm{pdf}_c(t)=\gamma_c\frac{\langle\psi(t)|\hat{c}^\dag\hat{c}|\psi(t)\rangle}{\langle\psi(t)|\psi(t)\rangle}$. The operator $\hat{a}$ represents the chance of photon leakage at rate $\gamma_a=\frac{\omega_a}{Q_\mathrm{opt}}$ with $Q_\mathrm{opt}$ the optical quality factor; $\hat{b}$ corresponds to a phonon leaking to the heat bath at rate $\gamma_b=\frac{\gamma_m (n_\textrm{th}+1)}{2}$, where $\gamma_m=\frac{2\Omega}{Q_\mathrm{mech}}$ with $Q_\mathrm{mech}$ the quality factor of the mechanical resonator (notice the different convention leading to a factor of 2 difference); lastly $\hat{b}^\dag$ corresponds to receiving a phonon from the bath at rate $\gamma_{b^\dag}=\frac{\gamma_m n_\textrm{th}}{2}$, where $n_\textrm{th}=\frac{k_b \tau}{\Omega}$ is the average number of phonons in the bath, $k_b$ is the Boltzman constant, and $\tau$ is the temperature of the bath. Solving for the dynamics and the probability densities of various jumps, as done in details in the interactive supplementary materials~\cite{supplementary_pluto} leads to:

\begin{enumerate}
    \item To zeroth order, no jump occurs.
    \item To first order, a photon-phonon pair is heralded. The probability of that event is $P_{a}=\int_0^T \mathrm{d}t\operatorname{pdf}_a(t)$.
    \item To second order, a photon-phonon pair is heralded and then followed by any other event, for an overall of probability $P_{a*}=1-\langle\psi(T)|\psi(T)\rangle$.
    \item Also to second order, a $b$ event at time $\tau$ is followed by an $a$ event, happening with $P_{ba}=\int_0^T \textrm{d}\tau\ \textrm{pdf}_b(\tau) \int_\tau^T\textrm{d}t\ \textrm{pdf}_{ba}(t)$.
    \item Similarly for $b^\dag$ followed by $a$ we have probability $P_{b^\dag a}$.
\end{enumerate}

Above, $T$ is the duration of the pump pulse. These are all the branches of the dynamics that have a chance of triggering a heralding event (to leading order). The total chance for heralding is $P=P_a + P_{ba} + P_{b^\dag a}$, while the fidelity of the heralded single phonon is $F=\frac{P_a - P_{a*}}{P_a + P_{ba} + P_{b^\dag a}}f_0$, where $f_0=\langle1|\rho_{a}|1\rangle$ is the fidelity of "good heralding" branch of the dynamics. $\rho_{a}$ is the density matrix for the state conditioned on only one $a$ event having happened during the pump pulse of duration $T$. \revisiontwo{The above trajectories can be seen pictorially in Fig. \ref{fig:protocol}c}.

After simplifying and taking into account that the decay of the optical cavity is much faster than the optomechanical interaction ($T_a=\gamma_a^{-1}\ll (\overline{a} g_{OM})^{-1}$), we obtain:
\begin{equation}
  P = 4\overline{a}^2g_\mathrm{om}^2T_aT,
  \label{prob}
\end{equation}
\begin{equation}
  1-F = 8\overline{a}^2g_\mathrm{om}^2T_aT + \frac{3}{4}\gamma_m T_a T \left( 3n_\mathrm{th}+1\right).   
  \label{fid}
\end{equation}

\begin{figure}
    \centering
    \includegraphics[width=\columnwidth]{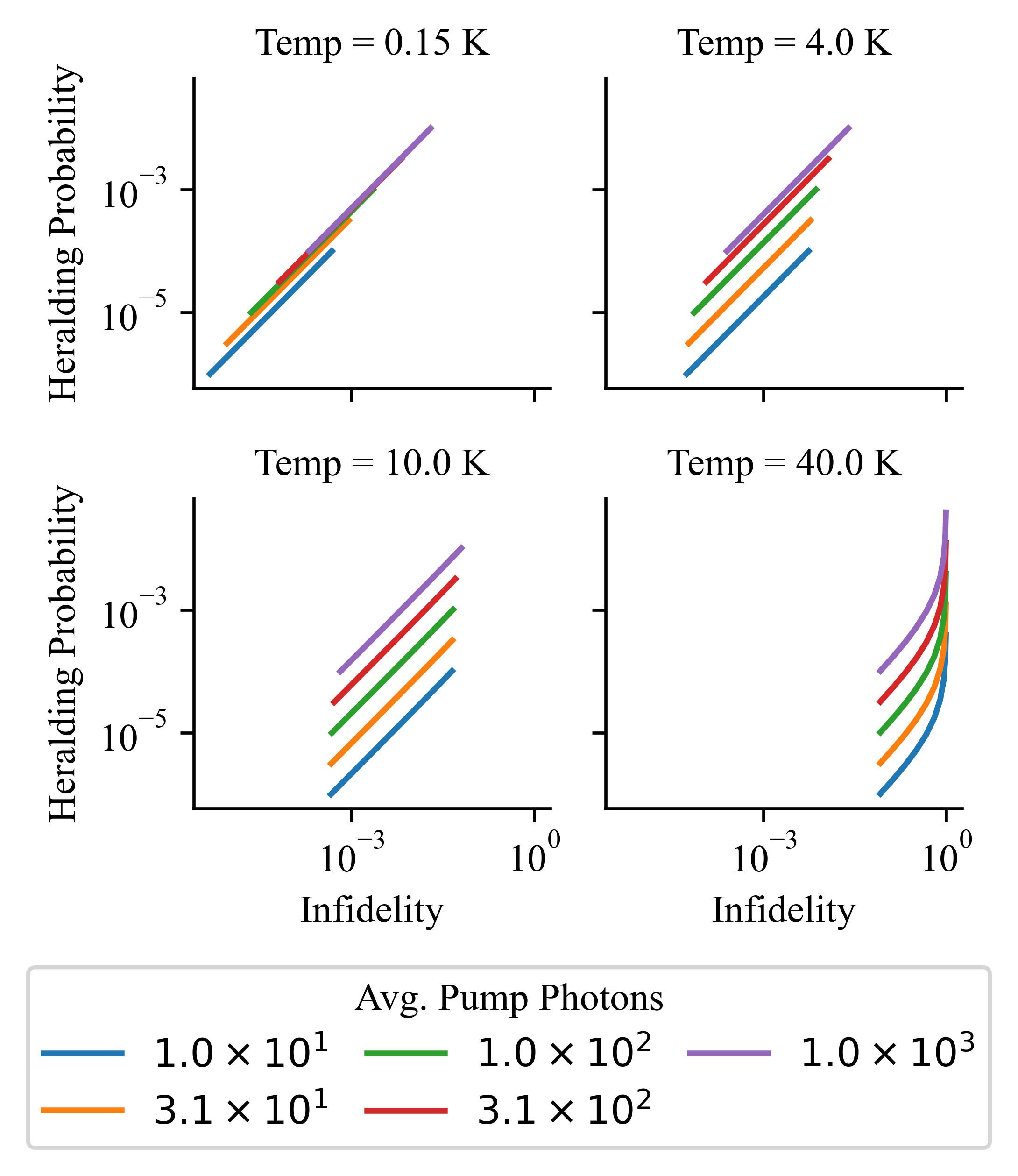}
    \caption{Heralding probability and single-phonon infidelities as a function of temperature (facet) and pump power (color), parameterized by pump pulse duration (each line spans $T=T_a$ to $T=10^3T_a$). Shorter pulses have lower probability and infidelity. However, the rate of heralding is independent of $T$ as the shorter the pulse (the higher the repetition rate), the lower the heralding probability for that attempt is. Therefore short pump pulses are preferable as that leads to lower infidelity. In this setup, at $\tau=40 K$, $\alpha=\sqrt{1000}$, and $T=T_a$, we can theoretically achieve rates of successful single-phonon heralding in the tens of kHz at infidelity lower than $10\%$. The performance is even better at lower temperatures. At around $4 K$ we see that the detrimental effects from the bath of the mechanical resonator become negligible compared to the infidelity due to multi-phonon excitations.}
    \label{fig:protocol_performance}
\end{figure}

Notice the term in the infidelity that scales exactly as the heralding probability: This is due to the $\mathcal(O)(\varepsilon^2)$ next-to-leading-order effect in the two-mode squeezing, leading to a proportionally larger chance of more-than-one excitations being heralded. There is also a second term, purely related to the detrimental effects of the thermal bath on the mechanical resonator. As long as $k_b\tau\ll Q_\mathrm{mech}\alpha^2g_\mathrm{om}^2$ we can neglect the bath heating term, however this can be difficult to quantify as $Q_\mathrm{mech}$ strongly depends on $\tau$. This transition between leading sources of infidelity can be seen in Fig.~\ref{fig:protocol_performance}.

These are the heralding probability and fidelity of a single phononic excitation. The heralding probability and fidelity for the complete entangling protocol, in which two nodes are pumped in parallel and the photon is looked for only after "path-information erasure" differ. To leading order, the probability $P_e=2P$ is twice as high as either node can produce a photon, and the infidelity scales the same.

For long term storage, we coherently swap the phononic excitation into the spin. The swap gate contributes an additional infidelity of $n_\mathrm{th}\gamma_m/g_\mathrm{sm}$ which is much lower than other sources of infidelity.

\begin{figure}
    \centering
    \includegraphics[width=\columnwidth]{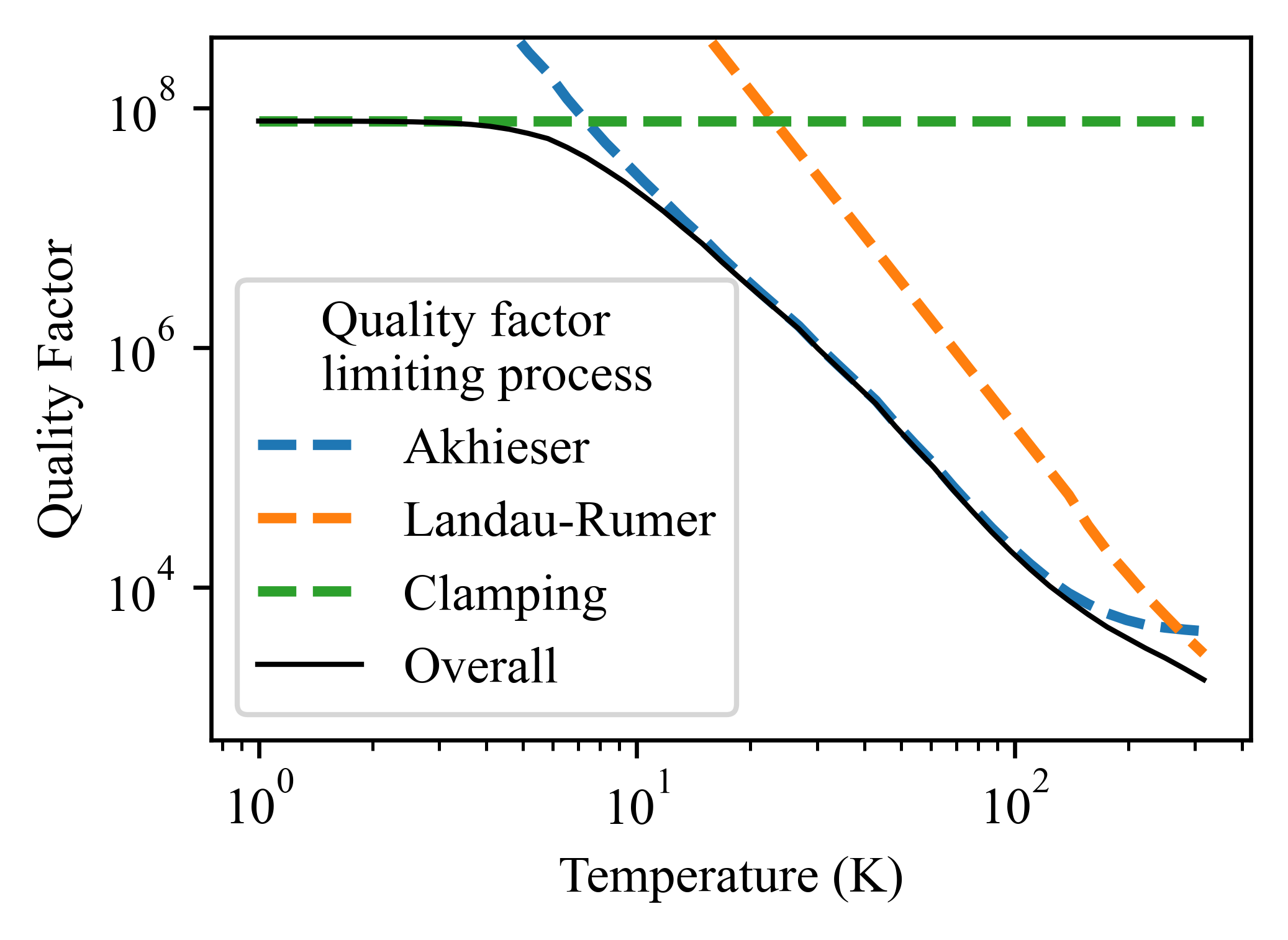}
    \caption{The processes limiting the quality factor of a mechanical resonator. At low temperature only clamping losses matter (green), but past a certain temperature Akhieser (blue) and Landau-Rumer (orange) processes dominate. These estimates depend on thermal properties of bulk diamond as reported in the literature. Thin-sheet diamond, as used in our devices, can have slightly differing properties.}
    \label{fig:akhieser}
\end{figure}

These results, given the design parameters of the previous section, are detailed in Fig.~\ref{fig:protocol_performance}. Of note is that $Q_\mathrm{mech}$ is very strongly dependent on the bath temperature due to scattering processes among the thermal phonons. At low temperatures, only clamping losses due to the design of the resonator are of importance, but as the temperature increases, Akhieser and then Landau-Rumer processes become important~\cite{ghaffari2013quantum,duwel2006engineering,kunal2011akhiezer,maris1971interaction}. The typical dependence for our design and material parameters can be seen in Fig.~\ref{fig:akhieser}. The Akhieser limited quality factor is $Q_A = \frac{1}{\Omega\tau}\frac{\rho c^4}{2\pi\gamma^2\kappa}$, where $\rho$ is density, $c$ is speed of light, $\gamma$ is the Gr\"uneisen coefficient, and $\kappa$ is the thermal conductivity. Only $\kappa$ depends strongly on temperature, and we use the values reported in~\cite{cvdbooklet,pohl1963applicability,berman1953thermal,pan1994diamond,barman2007temperature,graebner1992dominance}. At even higher temperature the Landau-Rumer processes dominate with $Q_{LR}=\frac{2\rho c^2}{\pi \gamma^2 C_v \tau}$, where $C_v$ is the diamond heat capacity as reported in~\cite{reeber1996thermal,corruccini1960specific}.

Thus, with our design we can theoretically achieve single-phonon generation at tens of kHz and infidelity lower than $10\%$ at temperature $\tau=40 K$, number of photons in the pump mode $\alpha^2=1000$, pump pulse duration $T=T_a$. At lower temperatures the performance significantly improves, giving limiting infidelities far below $1\%$.

\subsection{Quantum Monte Carlo verification and Design Feedback}

\begin{figure*}[h]
\includegraphics[width=\textwidth]{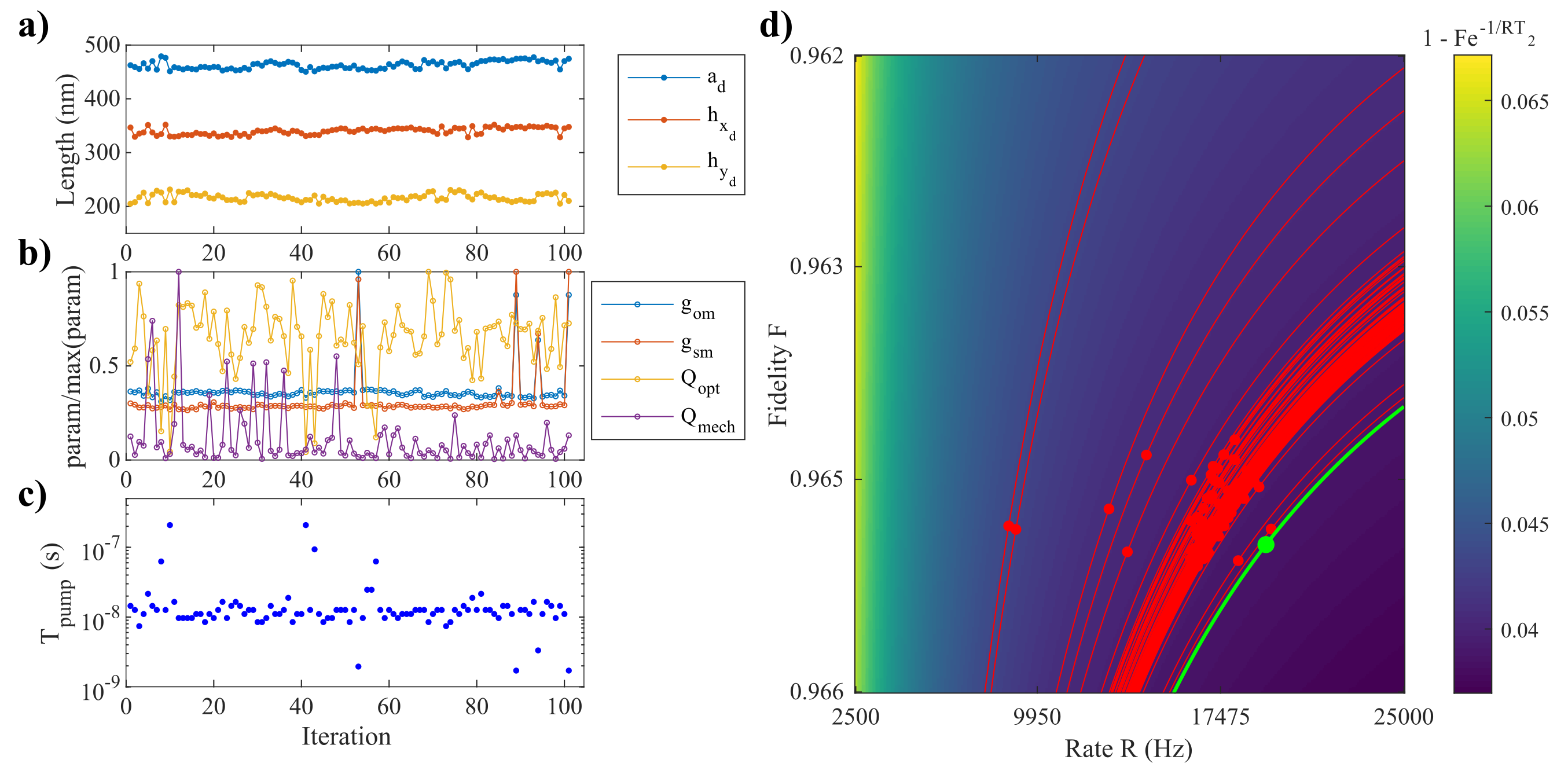}
\caption{Design feedback using FEM-QuTiP optimization. (a) shows the change in $a_d$, $h_{x_d}$, and $h_{y_d}$ at each Bayesian optimizer iteration, where the resulting performance parameter variation is shown in (b). For each iteration of \{$a_{d}, h_{x_{d}}, h_{y_{d}}$\}, optimizer individually selects an optimum pump time $T_{pump}$ which minimizes the cost function for that iteration and (c) shows the change in $T_{pump}$ by iteration. (d) shows the rate-fidelity tradeoff of the cost function at each iteration (red), with the optimized point highlighted (green). The red (green) lines represent curves in the rate-fidelity space with equal cost function to the iteration points.}
\label{fig:optimization_sweep}
\end{figure*}

Finally, we use the protocol above to feedback on the optomechanical crystal design parameters (namely, $a_{d}$, $h_{x_d}$, and $h_{y_d}$) (Fig.~\ref{fig:optimization_sweep}(a-b) to minimize the cost-function which is a function of fidelity and rate of heralding (see Supplementary). We verify the theoretical equations Eq. ~\ref{prob} and Eq.~\ref{fid} using a quantum monte-carlo  approach (see Supplementary) and proceed to use a COMSOL-to-Python (theory + quantum master equation) feedback loop to optimize the crystal design.

\revisiontwo{We select a bridge width $b$ of 60 nm, taking into account the nanofabrication considerations, and defect unit cell number $n_{d}$ as 6. With these parameters, we run a bayesian optimization over the free variables $h_{y_d}$, $h_{x_d}$, and $a_{d}$, using COMSOL finite element method (FEM) simulation to extract device parameters $\{\omega_a$, $\Omega$, $Q_{opt}$, $Q_{mech}$, $g_{om}$, $g_{sm}$\}. The COMSOL simulation is interfaced with python API using the MPh module, which for each iteration transfers the COMSOL results to qutip, where Eq. ~\ref{prob} and ~\ref{fid} estimate the phonon heralding and QuTiP quantum master equation simulates the phonon-to-spin swap. This then evaluates the heralding success rate and fidelity for the protocol described in Section 2.3 (see Supplement for more details).} Through this optimization, we arrive at a final design parameter set in Table \ref{optomech table}.

\begin{table*}[h]
\centering
\begin{tabular}[width=\textwidth]{c|c|c|c|c|c|c|c|c|c|c|c|c|c}
  \hline \hline Mat. & Emitters & a & a$_d$ & $h_x$ & $h_{x_d}$ & $h_y$ & $h_{y_d}$ & w & t & b & $Q_{opt}$ & $Q_{mech}$ & $g_{om}/2\pi$ (Hz) \\\hline
    Dmd & SiV$^-$ & 577.5 & 474.5 & 200 & 347.7 & 700 & 210.1 & 913.5 & 250 & 60 & 8.6e4 & 6.2e7 & 8.2e5 \\
    SiC & $V_{Si}^-$ & 480 & 299.6 & 235 & 230 & 600 & 340.4 & 750 & 250 & 60 & 2.8e3 & 2.0e6 & 1.9e6 \\
    Si & Si:B & 535 & 435 & 325 & 342.9 & 370 & 305 & 500 & 220 & 60 & 6.37e3 & 2.1e6 & 3.7e6\\
    \hline\hline
\end{tabular}
\caption{Spin-optomechanical cavities optimized by the FEM-QuTiP optimization in Section 2.4. The emitter used in the optimization protocol is listed next to each material. All parameters $\{a,a_d,h_x,h_{x_d},h_y,h_{y_d},w,t,b\}$ are listed in nm. For these optimizer runs, we assumed a temperature of 100 mK, and a $T_2$ of 13 ms\cite{PhysRevLett.119.223602}, 1 ms\cite{PhysRevA.109.022603} and 0.9 ms\cite{kobayashi2021engineering}, for SiV$^{-}$, $V_{Si}^-$ and Si:B respectively.}
\label{optomech table}
\end{table*}

Possible improvements to the protocol include (1) spectral and spatial multiplexing  (2) use of a Dicke state of multiple nearby color centers to enhance $g_\mathrm{sm}$ (3) use the nuclear registers for even longer storage times (4) entanglement purification with the nuclear registers which greatly increase the entanglement fidelity while only marginally decreasing the entanglement rate.

\section{Discussion}

In this paper, we bring the idea of a self-similar concentrator from photonic crystal devices \cite{choi2017self} to a 1D optomechanical crystal and explore the usage of the resulting cavity in spin-optomechanical interfacing. This system poses the advantages afforded by highly concentrated optical and mechanical modes: high strain in a central region while retaining optomechanical coupling in diamond relative to previously proposed and demonstrated devices~\cite{burek2016diamond}, and thus strong spin-phonon interactions. From FEM simulations, we demonstrate that this spin-optomechanical interface can \revisiontwo{achieve $820$ kHz single photon-phonon coupling alongside $32$ MHz} spin-phonon coupling to a Group IV spin. The strength of this spin-phonon interaction is such that we can effectively ignore losses incurred when swapping a quantum between a cavity phonon and the spin state.

We explore implementation of our interface in an optically heralded entanglement protocol \cite{duan2001long,krastanov2021opticallyheralded}. In this scheme, identical cavities are entangled via heralding, and the resulting entangled phonons are swapped into their respective coupled spins. This entanglement procedure completely circumvents standard issues related to spin-addressing, including the need to operate at the emitter's optical transition wavelength (we define the optical wavelength with a telecom photonic mode) and concerns related to spectral diffusion of emitters (we never optically excite the emitter). Additionally, this scheme places no strong requirements on the optical quality factors required by other works to accomplish spin-mechanical addressing~\cite{ji2020proposal,ghobadi2019progress}--instead operating with low optical $Q$s to increase the rate of heralding--and requires on-chip devices that are well within fabricable parameters. 

\begin{table*}
\centering
\begin{tabular}[width=\textwidth]{c|c|c|c|c}
  \hline \hline Defect & Material & $g_{{sm}_{proj}}/2\pi$ & $\sim T_1$ (@T) & $\sim T_2$(@T) \\\hline
    SiV$^-$ & Diamond & $\sim$\revisiontwo{32 MHz} & 0.1 ms @40mK\cite{becker2018all} &0.2 ms @40mK\cite{becker2018all}\\
    SnV$^-$ & Diamond & $\sim$26 MHz & 10 ms @3K\cite{trusheim2020transform} & 0.3 ms @1.7K\cite{debroux2021quantum} \\
    NV$^-$ & Diamond & $\sim$440 Hz &100 s @20K\cite{jarmola2012temperature} & 0.6 s @77K\cite{bar2013solid} \\
    Si:B & Si & \revisiontwo{$\sim$30 MHz} & 5 ms @25mK\cite{kobayashi2021engineering} & 0.9 ms @25mK\cite{kobayashi2021engineering} \\
    Si:P & Si & $\sim70$ MHz &0.3 s @7K\cite{tyryshkin2003electron} & 60 ms @7K\cite{tyryshkin2003electron} \\
    V$_{\text{Si}}$ & SiC & \revisiontwo{$\sim10$ MHz} & $10$ s @17K\cite{simin2017locking}& $20$ ms @17K\cite{simin2017locking}\\
    VV$^0_{\text{Si}}$ & SiC &  $\sim 2$ kHz & 8 ms @20K\cite{falk2013polytype} & 1 ms @20K\cite{christle2015isolated}\\
    \hline\hline
\end{tabular}
\caption{Spin defect candidates for optomechanical interfacing. The defects and their host materials, projected couplings $g_{sm_{proj}}$, and measured $T_1$ and $T_2$ at different operating temperatures are listed. The $g_{sm_{proj}}$ of the SnV$^-$, NV$^-$, V$_{\text{Si}}$, and VV$^0_{\text{Si}}$ were estimated by modifying parameters in the SiV$^-$ coupling formula \cite{trusheim2020transform,ovartchaiyapong2014dynamic,teissier2014strain,whiteley2019spin,udvarhelyi2020vibronic,udvarhelyi2018ab}. The B:Si and P:Si $g_{sm_{proj}}$ were estimated by substituting our $b = 60$ nm mode volume into the formulae in \cite{ruskov2013onchip} and \cite{soykal2011sound}, respectively. The defect naming conventions have been copied from references.}
\label{spin table}
\end{table*}

Our spin-optomechanical architecture applies to other material platforms besides diamond.
For example, silicon (Si) and silicon carbide (SiC) have been used for optomechanics \cite{eichenfield2009optomechanical,chan2012optimized, ren2020two, lu2015high,lu2019silicon,lu2020silicon} and have quantum emitters including carbon-based T-centers, phosphorus vacancies, and boron impurities \cite{bergeron2020silicon, ruskov2013onchip}. In particular, Si with B:Si acceptor impurities has been considered for operating spin-phonon coupled systems as an acoustic alternative to circuit-cavity QED \cite{ruskov2013onchip}. Here, we have shown that with an intentionally designed optomechanical cavity, one can achieve $g_{sm}$ much larger than previously proposed--which should be the case irrespective of the material, whether diamond, silicon, or another alternative--alongside respectable $g_{om}$, such that the full spin-optomechanical interface's performance can be evaluated (see Table ~\ref{spin table}). We have analyzed this interface assuming a SiV$^-$ spin, which has well-documented spin-strain parameters \cite{hepp2014electronic,meesala2018strain}; \revision{however the spin-dephasing time is highly limited above single-Kelvin temperatures due to electron-phonon dephasing \cite{jahnke2015electron}}. As such, future works may use this spin-optomechanical framework while selecting a suitable combination of material platform and temperature-robust, highly strain-tunable spin defect. \revision{The beauty of this platform is that, given sideband-resolved cooling of nanomechanical oscillators at a few Kelvin\cite{qiu2020laser} or $\sim 20$ K\cite{chan2011laser}, quantum operation of a solid-state spin would not be limited by optical lifetimes and instead enabled by state-of-the-art optomechanics.} The ability to separately engineer quantum memories and spin-photon interfaces, while retaining efficient interfacing between them even at moderate temperatures up to 40 K, will provide much-needed design freedom in applications from quantum networks to computing to sensing. \revisiontwo{Our COMSOL file, code for COMSOL-python optimizer, QuTiP simulation can be found in our GitHub repository\cite{panand2257}}.

\section{Acknowledgements}
The authors would like to thank Gerry Gilbert and Gen Clark for insightful comments on this research.
The Julia and Python open source communities provided invaluable research software. The hardware design was done in COMSOL. SK and HR are grateful for the funding provided by the MITRE Quantum Moonshot Program. HR acknowledges support from the NDSEG Fellowship and the NSF Center for Ultracold Atoms. PA acknowledges support from the Center for Quantum Networks. DE acknowledges support from NSF.  D.E. holds shares in Quantum Network Technologies. 

\bibliography{bib}
\clearpage

\onecolumn

\section{Supplementary Material}

\subsection{Equation Recap for Spin-Optomechanical Interface}

The spin-optomechanical crystal is governed by the coupling parameters $g_{om}$ and $g_{sm}$. The single photon-to-single phonon coupling between a photonic cavity mode and a mechanical resonant mode arises due to the cavity frequency shift induced by the acoustic displacement profile, normalized to the zero-point fluctuation governed by the resonator's effective mass $m_{eff}$: ~\cite{johnson2002perturbation,safavi2014optomechanical}
\begin{equation}
    g_\mathrm{om} = \frac{\partial\omega}{\partial\mathbf{q}}x_\mathrm{zpf}, \;
    x_\mathrm{zpf} = \sqrt{\frac{\hbar}{2 m_\mathrm{eff}\Omega}}, \;m_\mathrm{eff} = \frac{\int_{V}d\mathbf{r} \mathbf{Q}^*(\mathbf{r})\rho(\mathbf{r})\mathbf{Q}(\mathbf{r})}{\max(\abs{\mathbf{Q}(\mathbf{r})}^2)}.
\end{equation}
Here, $\mathbf{Q}(\mathbf{r})$ is the mechanical displacement profile and $\rho(\mathbf{r})$ is the density profile (either $\rho_\mathrm{diamond}$ or 0).

This consists of two explored effects: the moving boundary effect (shift due to moving vacuum-dielectric boundary conditions resulting from mechanical displacement) and the photoelastic effect (frequency shift due to the sum of strain-induced local refractive index changes in the crystal). The vacuum moving boundary coupling $g_{mb}$ can be written as~\cite{eichenfield2009optomechanical,johnson2002perturbation}

\begin{equation}
    \frac{g_\mathrm{mb}}{x_\mathrm{zpf}} = -\frac{\omega_\mathrm{a}}{2}\frac{\int_{S}\left(\mathbf{Q}(\mathbf{r})\cdot \mathbf{n}\right)\left(\Delta\boldsymbol{\varepsilon}\abs{e^\parallel}^2 - \Delta(\boldsymbol{\epsilon}^{-1})\abs{\overline{d}^\perp}^2\right)dA}{\max(\abs{\mathbf{Q}})\int\boldsymbol{\varepsilon}(\mathbf{r})\abs{\mathbf{e}(\mathbf{r})}^2d^3\mathbf{r}}.
    \label{gMB}
\end{equation}


The photoelastic coupling $g_\mathrm{pe}$ can be expressed as~\cite{safavi2014optomechanical}

\begin{align}
    \frac{g_\mathrm{pe}}{x_\mathrm{zpf}} =  -\frac{\omega_\mathrm{a}}{2}\frac{\int_V\mathbf{e}\cdot\boldsymbol{\delta\epsilon}\cdot\mathbf{e} d^3\mathbf{r}}{\max(\abs{\mathbf{Q}})\int_{V}\boldsymbol{\epsilon}(\mathbf{r})\abs{\mathbf{e}(\mathbf{r})}^2 d^3\mathbf{r}}. 
    \label{gPE}
\end{align}

Here, $\mathbf{e}(\mathbf{r})$ is the cavity electric field profile. Expanding the integrand in the numerator of \eqref{gPE}, we can write that as~\cite{chan2011laser}
\begin{align}
    \mathbf{e}\cdot\boldsymbol{\delta\epsilon}\cdot\mathbf{e} &= \mathbf{e}\cdot\left(\varepsilon^2\frac{\mathbf{p}\mathbf{S}}{\varepsilon_0}\right)\cdot\overline{e}\\
    &= \mathbf{e}\cdot\left(\epsilon_0 n^4 p_{ijkl}(\alpha)S_{kl}\right)\cdot\mathbf{e},
\end{align}

\revision{where $p$ is the photoelastic tensor of diamond \cite{lang2009strain,burek2016diamond}.} Here, we note that $p_{ijkl}$ is a function of the diamond crystallographic orientation relative to the device geometry, which runs along $\hat{x}$ in the $xy$-plane. Parametrized by $\alpha$ (the angle between the [100] crystal axis and the longitudinal axis of the nanobeam), the rotated $p_{ijkl}$ is given by~\cite{chan2011laser}
\begin{equation}
    p_{i j k l}(\alpha) = R(0,\alpha)_{iq}R(0,\alpha)_{jr}R(0,\alpha)_{ks}R(0,\alpha)_{lt} p_{qrst},
\end{equation}
where
\begin{equation}
    R(\theta, \phi) = \begin{bmatrix}
    \cos\phi & \sin\phi & 0 \\
    -\cos\theta\sin\phi & \cos\theta\cos\phi & -\sin\theta \\
    -\sin\theta\sin\phi & \sin\theta\cos\phi & \cos\phi
    \end{bmatrix}.
\end{equation}
In our simulations, we used $(p_{11}, p_{12}, p_{44}) = (-0.25, 0.043, -0.172)$~\cite{lang2009strain}.

The ultimate spin-phonon coupling is a function of the strain-induced $g_\mathrm{sm}$ profile (Fig.~\ref{fig: sweeps}(a)). We use $g_\mathrm{sm}$ to indicate the effective spin-orbital coupling resulting from a change in SiV$^{-}$ transition frequency as a function of displacement-induced strain~\cite{neuman2020phononic},
\begin{equation}
    g_\mathrm{sm}(\mathbf{r}) = d\frac{(\epsilon_{xx}(\mathbf{r}) - \epsilon_{yy}(\mathbf{r}))}{max{(\abs{\mathbf{Q}})}}x_\mathrm{zpf}.
\end{equation}
Here, $d \approx 1$ PHz/strain is the strain-susceptibility parameter describing the mixing of SiV$^-$ orbitals, and $\epsilon_{xx}$ and $\epsilon_{yy}$ describe the strain tensor components of the SiV$^-$.~\cite{hepp2014electronic,meesala2018strain,neuman2020phononic}. The SiV$^-$ $\{x,y,z\}$-axis is offset from the diamond $\{x_0,y_0,z_0\}$ axis by polar angle $\theta = \arcsin\sqrt{2/3}$ rad and azimuthal angle $\phi = \pi/4$ rad~\cite{hepp2014electronic}. So, to get $\epsilon_{xx}$ and $\epsilon_{yy}$ of the SiV$^-$ from crystal tensor components, we apply the rotation operation
\begin{equation}
    \epsilon_{i j k l} = R(\theta,\phi)_{iq}R(\theta,\phi)_{jr}R(\theta,\phi)_{ks}R(\theta,\phi)_{lt}\epsilon_0{}_{qrst}(\alpha).
\end{equation}
Note that $g_\mathrm{sm}$ varies by location in the cavity; as such, \revision{we have plotted the mode profile of $g_\mathrm{sm}$ in Fig.~\ref{fig: sweeps}(a)}. 
We find that $g_\mathrm{sm}$ is maximized at an angle $\alpha = 3\pi/4$ rad, with a maximum value \revisiontwo{$g_\mathrm{sm}/(2\pi) \approx 32$ MHz}, owing to phase matching between $\epsilon_{0_{yy}}$ and $(\epsilon_{0_{xy}} + \epsilon_{0_{yx}})$ terms.

\subsection{Simulations of Optomechanical Crystal}
The diamond optomechanical crystal was designed and simulated using the finite element method (FEM) in COMSOL Multiphysics 5.4. Simulations began with the analysis of a nanobeam unit cell. Mechanically, we simulate the eigenmodes of the unit cell with Floquet boundary condition defined by
\begin{equation}
    \mathbf{Q}(x) = \mathbf{Q}(x)e^{ik_x x},
\end{equation}
where $\mathbf{Q}(x)$ is the mechanical displacement profile at a given $x$ and $k_x \in \{0, \frac{\pi}{a}\}$. Similarly, we simulate the electromagnetic eigenmodes using the Floquet boundary equation
\begin{equation}
    \mathbf{e}(x) = \mathbf{e}(x)e^{i k_x x},
\end{equation}
where $\mathbf{e}(x)$ is the electric field at a given $x$. Bandstructures for these simulations are shown in Fig.~\ref{fig:bandstructures}c-d in the main text. After locating optical and mechanical bandgaps, the unit cells were varied by changing unit cell parameters $\{a, h_x, h_y\}$ to their "defect unit cell" values of $\{a_d, h_{x_d}, h_{y_d}\}$ to re-simulate mechanical and optical bandstructures at the $\Gamma$ point (for the mechanical breathing mode) and the $\text{X}$ point (for the electromagnetic confined mode).
Finally, the central two unit cells were modified to add the concentrating taper. A rectangular section of length $2 a_d$ and height $2 h_{y_d}$ was subtracted from the center of the crystal, connecting the two central ellipses. Next, the central taper was filled according to the intersection of the rectangular top and bottom lines with two hyperbolic curves following the (right hand side) equation (mirrored on the left-hand side)
\begin{equation}
    x(y) = \frac{c_1}{c_2} \sqrt{c_2^2 + y^2}, c_1 = \frac{b}{2}, c_2 = \frac{b h_{y_d}}{2 a_d}.
\end{equation}

This resulted in the geometry shown in Fig.~\ref{fig:bandstructures}a. The completed nanostructure was then simulated using the Solid Mechanics and Electromagnetic Waves, Frequency Domain (ewfd) physics modules and Eigenmode solver in COMSOL. The introduction of a central taper necessitated further modifying of the defect unit cell parameters, so $\{a_d, h_{x_d}. h_{y_d}\}$ were varied slightly to induce optical and acoustic modes that were near the center of the optical and acoustic band gaps, respectively, of the mirror unit cells.



From these simulations, we find that $Q_{mech}$ will likely not limit the overall mechanical quality factor. $Q_{opt}$ can likely be tuned for each $b$ according to radiation cancellation \cite{choi2017self}, but for our heralded entanglement protocol, $Q_{opt}$ primarily dictates the detection rate of a photon-phonon pair generation event. This leakage rate should be greater than the rate of phonon decay (i.e. $\kappa_{opt} \equiv \frac{\omega_a}{Q_{opt}} > \frac{\Omega}{Q_{mech}} \equiv \kappa_{mech}$) to ensure the acoustic phonon is not lost by the time the accompanying photon is detected. Hence, it is not important (nor necessarily favorable) for us to optimize $Q_{opt}$ individually in this paper; we optimize using the cost function in the main text instead.

\subsection{Calculation of Spin-Phonon Coupling}
Spin-phonon coupling was calculated using FEM simulations in COMSOL. The Euler angle $\alpha$ representing the in-plane rotation of the diamond crystal orientation relative to the $x$-axis of the nanobeam was swept as $0^{\circ} \leq \alpha \leq 180^{\circ}$. For the subsequent calculations of $g_{sm}$, we assume that the diamond crystal $z$-axis is oriented along the high-symmetry axis of the defect $[111]$, $x$ along $[\bar{1}\bar{1}2]$, and $y$ along $[\bar{1}10]$, such that the SiV$^-$ experiences \cite{neuman2020phononic}
\begin{equation}
    \epsilon_{xx} - \epsilon_{yy} = \frac{1}{3}\Big(-\epsilon_{11} - \epsilon_{22} + 2 \epsilon_{33} + 2(\epsilon_{12} + \epsilon_{21}) - (\epsilon_{13} - \epsilon_{31}) - (\epsilon_{23} + \epsilon_{32})\Big).
\end{equation}

\begin{figure}[h]
    \centering
    \includegraphics[width=\textwidth]{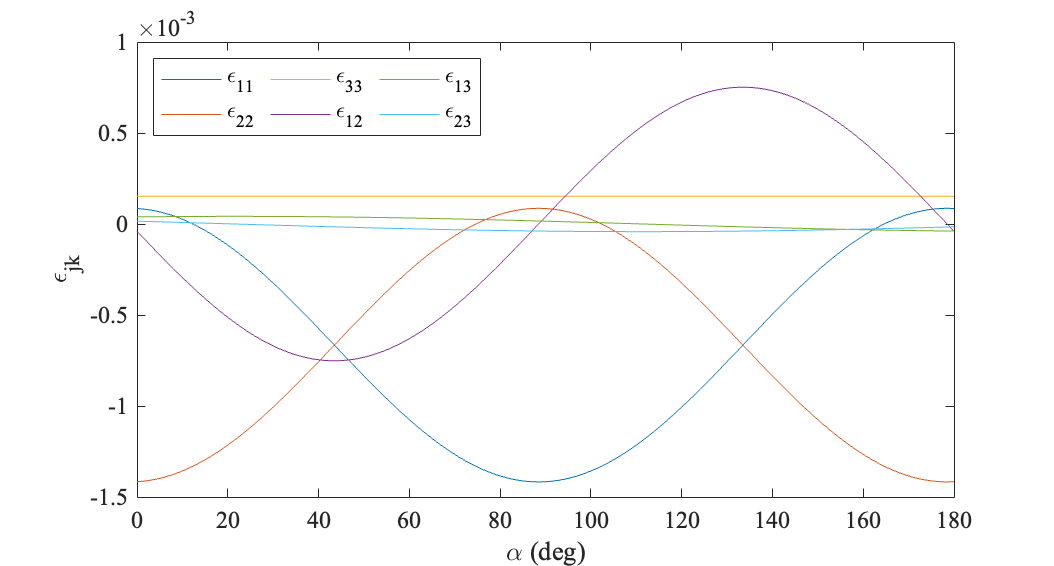}
    \caption{Sweep of the diamond crystal orientation strain tensor elements $\epsilon_{jk}$ with respect to $\alpha$. These tensor components were calculated at the middle of the top-right edge of the central taper for the \revisiontwo{5.39 GHz} acoustic mode ($b = 60$ nm).}
    \label{fig:eJK sweep}
\end{figure}

The $g_{sm}$ for each $b$ was calculated by taking the coupling at the middle of the right-edge of the central taper at ten slices from $z = t_d$ to $z = 0$. This was done to reduce numerical noise in the FEM simulation that resulted from extremely small elements experiencing dramatic deformation without increasing the mesh density to untenable levels.
\begin{figure}[h]
    \centering
    \includegraphics[width=\textwidth]{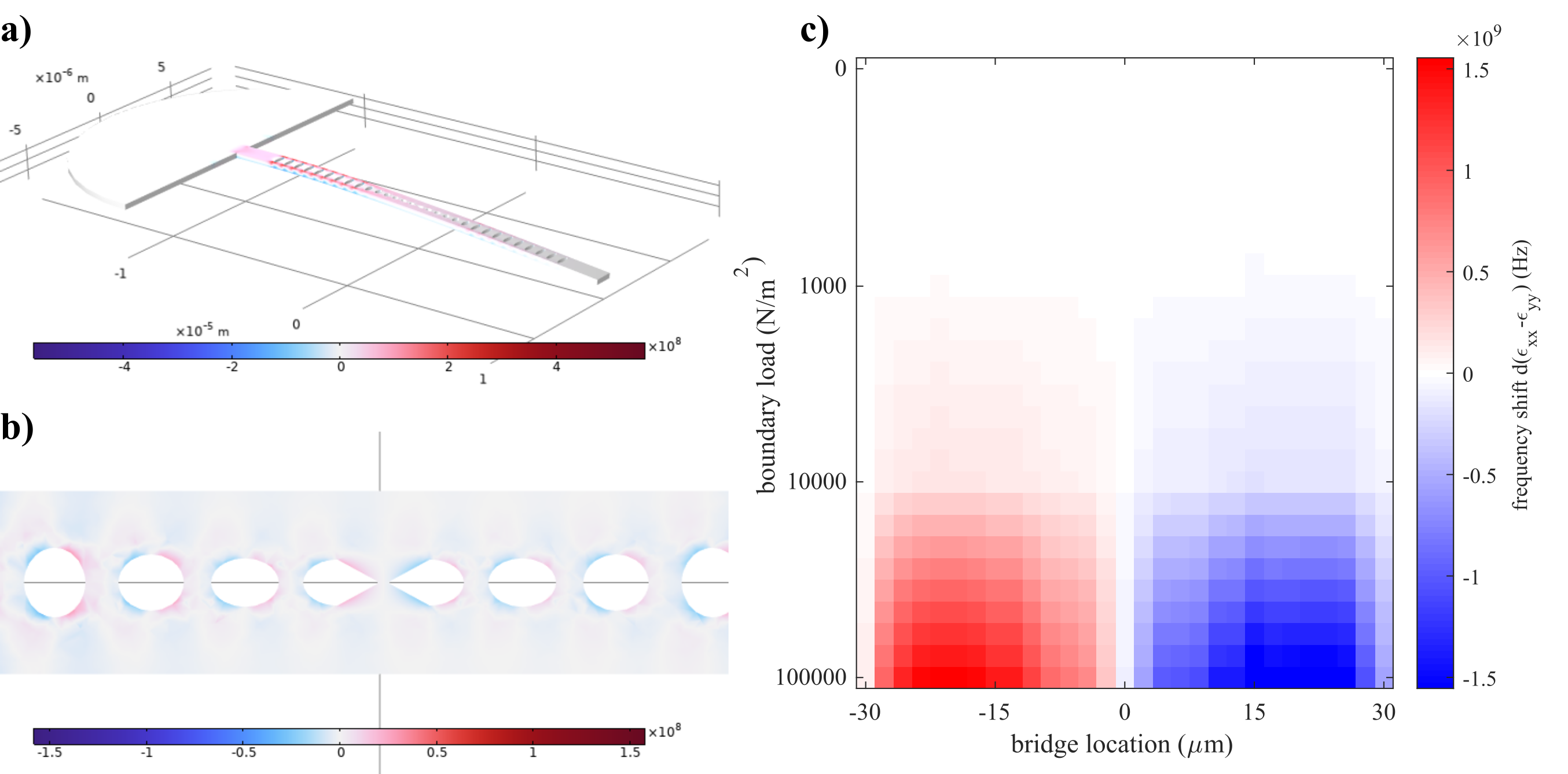}
    \caption{Plot of (a) frequency shift profile in a singly clamped cantilever with boundary load at the unclamped end. (b) Slice of frequency shift in the center thickness of the diamond OMC when applying said boundary load of 100 $N/m^2$. (c) the frequency shift profile of the central bridge as a function of different boundary loads. This frequency shift profile indicates the feasibility of strain actuating the spin to detune it from the phononic mode when turning off the spin-phonon interaction if it is desirable to detune the spin after an entanglement protocol succeeds.}
    \label{fig:DC tuning}
\end{figure}

We note here that there is a distinction between the strain-orbital coupling $d = \epsilon_{xx} - \epsilon_{yy}$ and the spin-strain coupling $g_{sm}$ dependent on the applied magnetic field. However, based on Appendix A in \cite{raniwala2022spin}, the spin-strain coupling can become comparable to the strain-orbit coupling when the vector magnetic fields reach values of $\sim 2$ T. This is achievable with commercially available vector magnets that can be added to cryostats.

\subsection{FEM-QuTiP Optimization and Quantum Monte-Carlo Simulations}
\begin{figure}
    \centering
    \includegraphics[width=\textwidth]{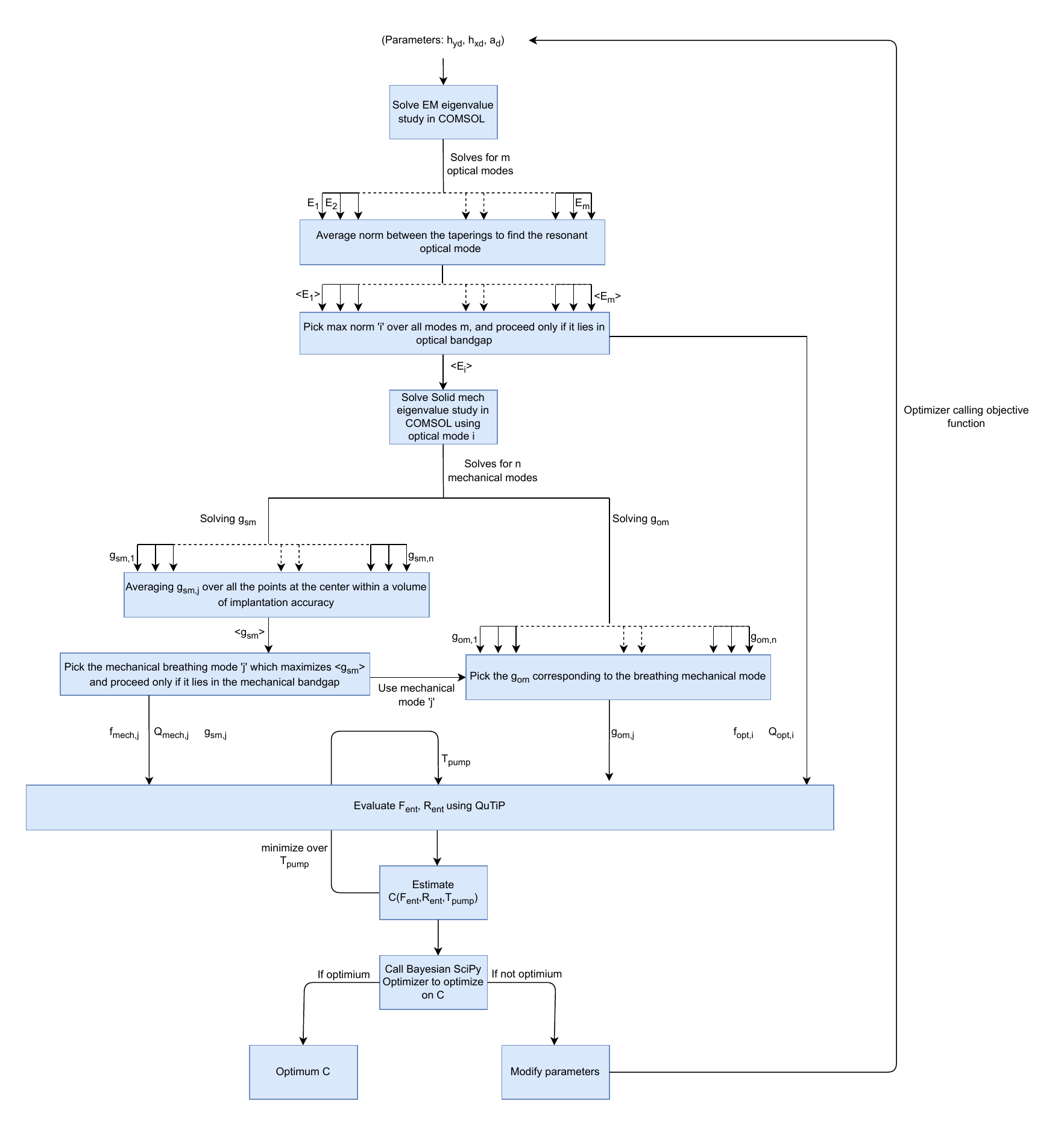}
    \caption{Flowchart for FEM and QuTiP optimization of the optomechanical crystal.}
    \label{fig:optimizer_flowchart}
\end{figure}

FEM-and-QuTiP optimization follows the flowchart in Fig.~\ref{fig:optimizer_flowchart}. First, we run electromagnetic (EM) FEM simulation using COMSOL-Python API to determine the optical mode properties of the optomechanical crystal. The optical mode describing the fundamental mode is filtered from the EM FEM results by identifying the optical mode with maximal electromagnetic field energy density averaged in the region between the taperings. We extract the optical frequency and quality factor ($f_{opt}$ and $Q_{opt}$, respectively) for this mode from the EM FEM result. Next, we run a structural mechanics (SM) FEM simulation through the COMSOL-Python API to determine the mechanical modes of the system. For each of the eigenmodes COMSOL solves for, we average the spin-mechanical coupling ($g_{sm}$) over a cube of side length 30 nm at the center of the taper to account for the implantation accuracy of the color center. We identify the breathing mode by selecting the mechanical mode which maximizes the absolute value of the averaged $g_{sm}$. We extract the values for the mechanical mode frequency, mechanical quality factor and the opto-mechanical coupling ($f_{mech}$, $Q_{mech}$, and $g_{om}$, respectively) corresponding to the breathing mode, from the SM FEM result. Through manual verification, we confirmed that this method works nicely for selecting both the fundamental optical mode as well as the breathing mechanical mode. The above obtained values are then fed into the QuTiP code which performs the following operations:
\begin{enumerate}
    \item Use FEM results and run QuTiP's master equation solver to simulate the fidelity corresponding to the cooling step 
    \item Then analytically calculate the phonon-photon pair heralding success probability $P_{herald}$, the fidelity of the heralded phonon state $F_{herald}$, and rate of heralding $R_{herald}$.
    \item Perform a spin-phonon rabi oscillation simulation using QuTiP's master equation solver to calculate the fidelity of the final spin state, $F_{swap}$.
    \item Estimate the entanglement rate $R_{ent}$ and fidelity $F_{ent}$ from $R_{herald}$ and $F_{herald}$.
    \item Select the optimal pump time $T_{pump}$ which minimizes the cost function $C(F,R,T_{2}) = 1 - Fe^{-1/RT_{2}}$.
    \item Report this optimum cost function to the optimizer for this iteration.
\end{enumerate}
Finally, the above FEM and QuTiP steps are fed into a built-in SciPy Bayesian optimizer for up to 100 iterations to minimize $C(F,R,T_{2})$ over the optomechanical crystal parameters $\{h_{x_d}, h_{y_d}, a_d\}$. The optimizer results are shown in Fig.~\ref{fig:optimization_sweep} and Table~\ref{optomech table}.

\begin{figure}
    \centering
    \includegraphics[width=0.85\textwidth]{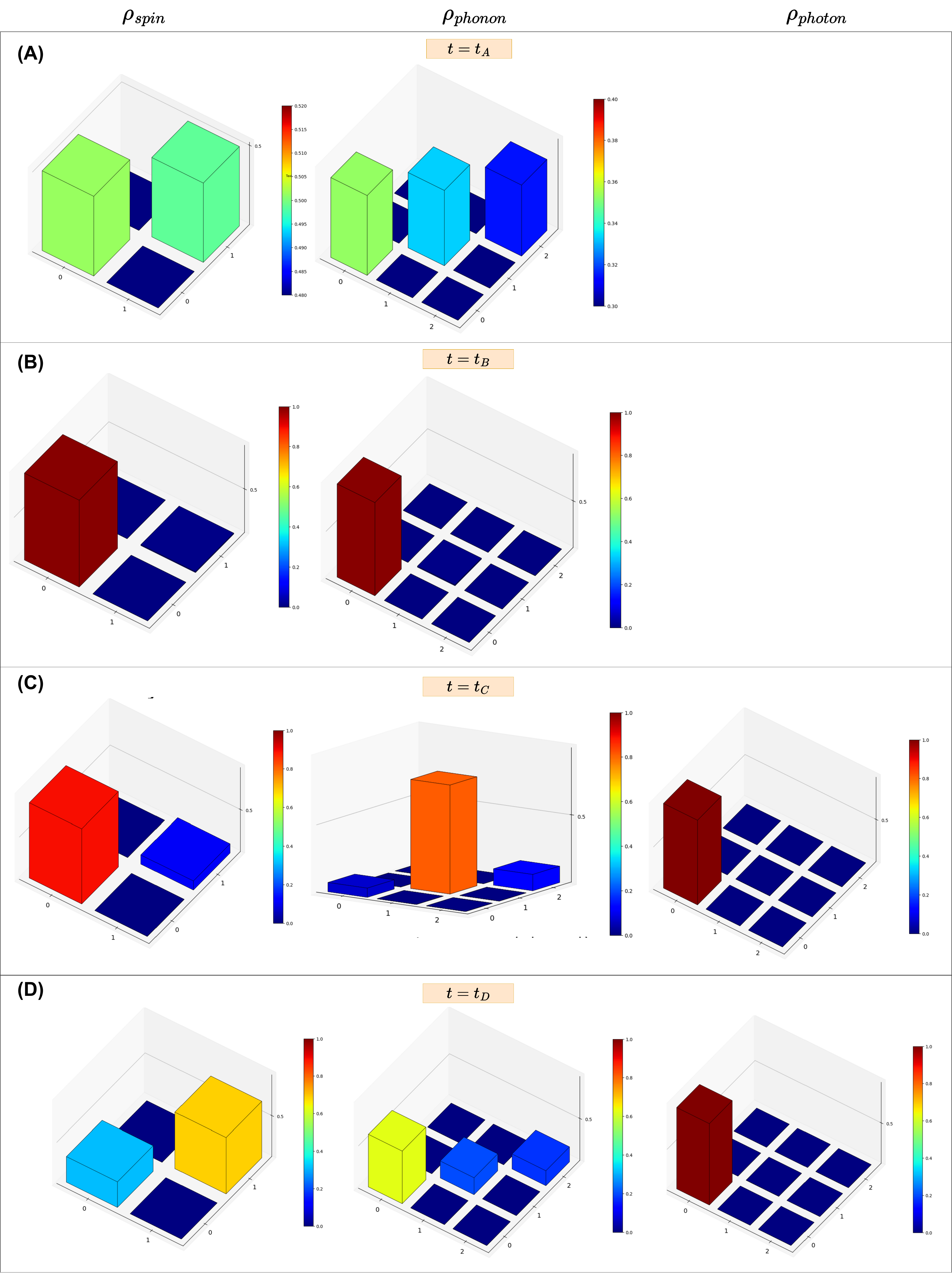}
    \caption{Simulated density matrices for the four time  steps of the protocol in the Fig.~\ref{fig:protocol}. We assume phonon and photon to be in Fock space of dimension 3. (A) At $t=t_{A}$, we start with the spin and phonon in their respective thermal states. After the cooling step, (B) at $t=t_{B}$ we see that most of the spin and phonon population is in their ground states. After this step, (C) at $t=t_C$ we start the blue-detuned squeezing of optomechanical modes, which would mean that we should have phonon in $|1\rangle$ state, and photon in $|0\rangle$, as the optical photon leaked into the waveguide from its $|1\rangle$ state, meaning in phonon-photon basis, we see $|00\rangle\rightarrow|11\rangle\rightarrow|10\rangle$, which is visible in the simulation. Finally, at (D) $t=t_{D}$, spin and phonon is swapped, and we see spin in excited state and phonon in $|0\rangle$ state, with some infidelities due to the bad trajectories as discussed in Fig.~\ref{fig:protocol}c.}      
    \label{fig:density-matrices}
\end{figure}

As a summary, the optimization routine is implementing the following task:
\begin{equation}
    \min_{\{h_{y_{d}},h_{x_{d}},a_{d},T_{pump}\}}C(F,R,T_{2})
\end{equation}
The cost function can be further tailored based on the quantum protocol that needs to be implemented, and our work demonstrates a tool for protocol specific device optimization\cite{doi:10.1126/sciadv.adg6685,doi:10.1021/acsphotonics.1c01651}. This motivates our future work towards Artificial Intelligence (AI) based co-design of quantum nodes and using Reinforcement Learning (RL) based strategies for quantum computing tasks\cite{li2024dynamic, 10.1117/12.3023224}.
Additionally, we also simulated the full protocol using QuTiP quantum Monte-Carlo\cite{molmer1993monte} method, which includes cooling, heralding, conditional swap and then finally reset. The reason we chose quantum Monte-Carlo method for simulating the Hamiltonian is because there is a conditional swap involved, which performs the electron-phonon swap on the condition if there is a photon click or not. This can be easily simulated by the quantum trajectory approach, and Fig.\ref{fig:protocol}c shows the map of all possible trajectories. Using quantum Monte-Carlo, we can estimate the statistics of all these sub-trajectories, and further we can perform a conditional swap only on the click trajectories. Fig.~\ref{fig:density-matrices} shows the simulated density matrix plots for the four time slices. The quantum Monte-Carlo is performed for 1000 trajectories, with an added nuance that we increased our $n_{pump}$ such that 1000 trajectories is sufficient to give us a decent statistics about heralding. Because in cases when $n_{pump}$ is low, the $p_{good-traj}$ becomes lesser than 0.001, which means that we would need trajectories much larger than $\sim1/p_{good-traj}$ to obtain decent statistics which puts constraint on the simulation time. Therefore, we decided to simulate the quantum Monte-Carlo with 1000 trajectories and higher value of $n_{pump}$ during the process of heralding. 

We also simulate the DC strain tuning of the spin in a singly clamped version of the spin-optomechanical interface in Fig. \ref{fig:DC tuning}. While the simulation features a simple boundary load at the unclamped end of the crystal, such a tuning mechanism can be achieved using electromechanical tuning or other techniques on-chip \cite{sohn2018controlling,clark2024nanoelectromechanical}

\begin{figure}
    \centering
    \includegraphics[width=\textwidth]{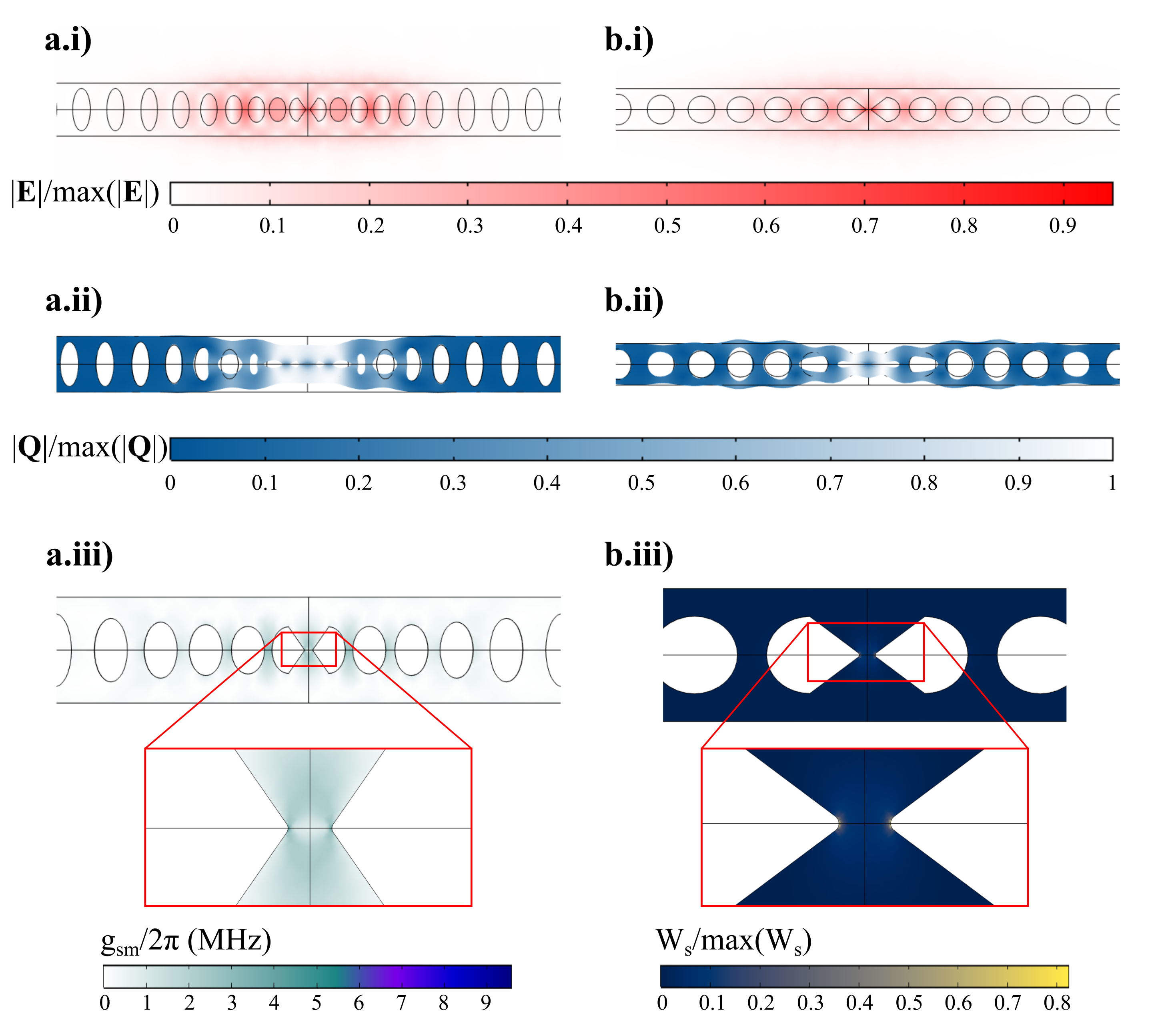}
    \caption{(a) Si optomechanical crystal with parameters $(a, a_d, h_x, h_{x_d}, h_y, h_{y_d}, w, t, b)$ = \revisiontwo{(535,432.96,325,323.42,370,289.47,500,220,60) [nm]} and (b) SiC optomechanical crystal with parameters $(a, a_d, h_x, h_{x_d}, h_y, h_{y_d}, w, t, b)$ = \revisiontwo{(480,285,235,215.15,600,330,750,250,60)} [nm]. (i) and (ii) show the mechanical and optical mode profile of each crystal, respectively. (a.iii) shows the predicted spin-mechanical coupling of a V$_\text{Si}$$^-$ vacancy to the mechanical mode, and (a.iii) shows the strain energy density of the Si crystal, which gives an estimate of spin-mechanical coupling to the Si mechanical mode.}
    \label{fig:Si cavity}
\end{figure}

\revision{\subsection{Analogous Spin-Optomechanical Interfaces in Silicon and Silicon Carbide}

Here, we demonstrate that this ultrasmall optical and mechanical mode volume cavity can, in principle, be achieved in different material systems featuring spins. Fig.~\ref{fig:Si cavity}(a) shows an optomechanical cavity with central taper in silicon \revisiontwo{($\omega_\mathrm{a}/2\pi = 188.6 \;\text{THz},\;\Omega = 4.07 \;\text{GHz}$), simulated to achieve a optomechanical coupling rate $g_{om}/2\pi \sim$0.91 MHz}  and featuring optical and mechanical mode volumes of \revisiontwo{$V_{\text{opt}}/\lambda^3\sim1.6 \times 10^{-3}$ and $V_{\text{mech}}/\lambda_{s}^3 \sim 4.75\times 10^{-4}$ $(V_{\text{mech}}/\lambda_{p}^3 \sim 1.4\times 10^{-4})$}, respectively. An ultrasmall mode volume optomechanical crystal in Si can be used to implement our spin-interfacing protocol with spin degrees of freedom in emerging color centers in Si, such as the metastable spin state of the G-center \cite{udvarhelyi2021identification,lee1982optical} or the ground-state spin of the T-center \cite{higginbottom2022optical}.

Fig.~\ref{fig:Si cavity}(b) shows a similar cavity with central taper in SiC \revisiontwo{($\omega_\mathrm{a}/2\pi = 199.1 \;\text{THz},\;\Omega = 6.37 \;\text{GHz}$), simulated to achieve a optomechanical coupling rate $g_{om}/2\pi \sim$0.69 MHz}  and featuring optical and mechanical mode volumes of \revisiontwo{$V_{\text{opt}}/\lambda^3\sim4.8 \times 10^{-3}$ and $V_{\text{mech}}/\lambda_{s}^3 \sim 4.82\times 10^{-4}$ $(V_{\text{mech}}/\lambda_{p}^3 \sim 1.43\times 10^{-4})$}, respectively.}

\end{document}